%% file: qcd-fenosoa.tex
\documentclass[final,3p,times,twocolumn]{elsarticle}
 \biboptions{comma,sort&compress}
\usepackage{here}
 \usepackage{graphicx}
  \usepackage{epsfig}
\def\nin{\noindent}
\def\beq{\begin{equation}}
\def\eeq{\end{equation}}
\def\bea{\begin{eqnarray}}
\def\eea{\end{eqnarray}}
\def\nnb{\nonumber}
\def\la{\langle}
\def\ra{\rangle}
\def\ga{\left(}
\def\dr{\right)}
\def\lrar              {\Longrightarrow}
\journal{Nuc. Phys. (Proc. Suppl.)}

\begin{document}

\begin{frontmatter}



\title{$\bar D^*D^*_0$ and $\bar B^*B^*_0~(1^{--})$ molecules at N2LO from QSSR$^*$}

\author[label1]{F. Fanomezana\corref{cor2}}
  \address[label1]{Institute of High-Energy Physics of Madagascar(iHEPMAD),University of Antananarivo, Madagascar 
}
\cortext[cor1]{Talk given at  QCD 14 (29 june - 3 july, Montpellier - France).}
\cortext[cor2]{Speaker.}
\ead{fanfenos@yahoo.fr} 
\author[label2]{S. Narison}
  \address[label2]{Laboratoire
Univers et Particules (LUPM), CNRS-IN2P3 \& Universit\'e
de Montpellier II, 
\\
Case 070, Place Eug\`ene
Bataillon, 34095 - Montpellier Cedex 05, France.}
\ead{snarison@yahoo.fr}
 \author[label1]{Andry Rabemananjara}  
 
\ead{achris\_01@yahoo.fr}
%\author{}

%\address{}

\begin{abstract}
\noindent
We estimate  the $\bar D^*D^*_0$ and $\bar B^*B^*_0~(1^{--})$ molecules masses and couplings using QCD spectral sum rules (QSSR) known perturbatively to N2LO of PT series and including the contributions of non-perturabtive condensates up to the dimension-eight. Our results improve earlier LO results obtained from QSSR in the current literature. We obtain $M_{bar D^*D^*_0}=5244(228)$ MeV which is heavier than the experimental candidates $Y(4260),~ Y(4360), ~Y(4660)$ suggesting that they cannot be pure molecule states. We predict $M_{\bar B^*B^*_0}$	= 11920(159) MeV to be tested in $B-$factory experiments. 
% approach allows a more precise value due to expansion in N2LO.
\end{abstract}


\end{frontmatter}

% \linenumbers

\section{Introduction}
%\label{}
\nin
%%%%%%%%%%%%
Motivated by the recently observed $1^{--}$ states Y(4260), Y(4360),  Y(4660) from their decays to $J/\psi\pi\pi$ decays \cite{BELLE} often interpreted as molecules or four-quark states \cite{REVMOL}, we improve our previous results \cite{10tetra} obtained to lowest order (LO) of PT series from the QCD spectral  sum rules \cite{SVZ,SNB} by evaluating the mass and coupling of the $\bar D^*D^*_0$ and $\bar B^*B^*_0~(1^{--})$ molecules at N2LO of PT QCD and including up to dimension 8 condensates. 
%%%%%%%%%%%%%%%%%%%%%%%%
\vspace*{-0.25cm}
\section{QCD analysis of spin one molecule}
\vspace*{-0.25cm}
%\label{}
\nin
%%%%%%%%%%%%
   \subsubsection*{$\bullet$ Current and two-point fonction}
   \nin
%%%%%%%%%%%%
The two-point correlation function associated to the $\bar D^*D^*_0$ and $\bar B^*B^*_0~(1^{--})$ molecule current  is defined as:
\bea
\Pi^{\mu\nu}_{mol}(q)&=&i\int d^4x ~e^{iq.x}\la 0
|TJ^\mu(x){J^\nu}^\dagger(0)
|0\ra\nnb\\
&=&-(q^2 g^{\mu\nu}-{q^\mu q^\nu})\Pi_{mol}(q^2)\nnb\\
&&+q^\mu q^\nu\Pi_{mol}^{(0)}(q^2)~,
\label{2po}
\eea
with the current:
\bea
 J^\mu &\equiv&(\bar Q\gamma^\mu q)(\bar q  Q )~,\\
 \nonumber Q &\equiv&  c, b\ ~~ {\rm and}~~ \ q \equiv u, d~,
 \eea
 where $\Pi_{mol}$ and  $\Pi_{mol}^{(0)}$ are respectively associated to the spin 1 and 0 molecule states. 
Due to its analyticity property, the correlation function $\Pi_{mol}(q^2)$ , obeys the dispersion relation:
\bea
\Pi_{mol}(q^2)=\frac{1}{\Pi}\int_{4m_c^2}^\infty ds \frac{\textrm{Im} \Pi_{mol}(s)}{s-q^2-i\epsilon}+...
\eea
where Im$\Pi_{mol}(s)$ are the spectral functions. 
   \subsubsection*{$\bullet$ Laplace sum rule (LSR)}
   \nin
%%%%%%%%%%%%
The Laplace SVZ sum rules are improvement of 
the previous dispersion relation which becomes after the (inverse) Laplace transform:
\beq
{\cal L}(\tau)=\int_{4m^2_Q}^{\infty} dt~ e^{-t \tau}  \frac{1} {\pi} {\rm Im} \Pi^{OPE}_{mol}(t)~.
\eeq
Parametrizing the spectral function by one resonance plus a QCD continuum, the lowest resonance mass $M_H$ and coupling $f_H$ normalized as:
\bea
\la 0|J^\mu|H\ra=f_H M_H^4 \epsilon^\mu~,
\label{eq:decay}
\eea
 can be extracted from the previous Laplace sum rules (LSR) as\,\cite{SVZ,SNB}:
\beq
M^2_H=\frac{\int_{4m^2_Q}^{t_c} dt~ t ~ e^{-t \tau}  \frac{1} {\pi} {\rm Im} \Pi^{OPE}_{mol}(t)}{\int_{4m^2_Q}^{t_c} dt~ e^{-t \tau}  \frac{1} {\pi} {\rm Im} \Pi^{OPE}_{mol}(t)}
\label{mass}
\eeq
and
\beq
f^2_H=\frac{\int_{4m^2_Q}^{t_c} dt~ e^{-t \tau}  \frac{1} {\pi} {\rm Im} \Pi^{OPE}_{mol}(t)}{e^{-\tau M^2_H}M^8_H}~,
\label{coupling}
\eeq
where $M_Q$ is the heavy quark mass, $\tau$ the sum rule parameter and $t_c$ the continuum threshold.
%%%%%%%%%%%%%%%%%%%%%%%%%%
\vspace*{-0.1cm}
 \subsubsection*{$\bullet$ The QCD two-point function at N2LO}
 \nin
For evaluating the perturbative part at NLO and N2LO, we assume a factorization of the bilinear currents and do a convolution of the corresponding  scalar Im$\Pi^{(0)}(s_1)$ and vector Im$\Pi^{(1)}(s_2)$ correlators \cite{PICH,SNPIVO}:
\bea
&&\frac{1}{\pi}{\rm Im}\Pi_{mol}^{(1)}(t)=\theta(t-4M^2_Q)\left(\frac{1}{4\pi}\right)^2 t^2 \int^{(\sqrt{t}-M_Q)^2}_{M_Q^2}\hspace*{-1cm} dt_1\times\nonumber\\ 
&& \int^{(\sqrt{t}-\sqrt{t_1})^2}_{M_Q^2} \hspace*{-1cm} dt_2 ~ \lambda^{3/2} \frac{1}{\pi}{\rm Im}\Pi^{(1)}(t_1) \frac{1}{\pi}{\rm Im}\Pi^{(0)}(t_2) 
\label{conv}
\eea   
with the phase space factor:
\bea
\lambda=\left(1-\frac{(\sqrt{t}-\sqrt{t_1})^2}{t}\right)\left(1-\frac{(\sqrt{t_1}+\sqrt{t_2})^2}{t}\right)~.
\eea   
Their QCD expressions are known in the literature \cite{SNFB12,SNFBST14,GENERALIS,CHET}.
We shall use the relation between the on-shell $M_Q$ and the running mass $\bar m_Q(\nu)$ to transform the spectral function into the $\overline{MS}$-scheme \cite{SPEC1,SPEC2}: 
\bea
M_Q &=& \overline{m}_Q(\nu)\Bigg{[}
1+{4\over 3} a_s+ (16.2163 -1.0414 n_l)a_s^2\nnb\\
&&+\ln{\ga\nu\over M_Q\dr^2} \ga a_s+(8.8472 -0.3611 n_l) a_s^2\dr\nnb\\
&&+\ln^2{\ga\nu\over M_Q\dr^2} \ga 1.7917 -0.0833 n_l\dr a_s^2\Bigg{]},
\label{eq:pole}
\eea
where $n_l=n_f-1$ is the number of light flavours and $a_s(\nu)=\alpha_s(\nu)/\pi$ at the scale $\nu$. 
The QCD  expressions of the non-perturbative part  of the $1^{--}$ molecule spectral functions used here have been  computed up to dimension $d=8$  in e.g. Ref. \cite{AlbNie1}.

%%%%%%%%%%%%%%%%%%%%%%%%%%%%%%%%%%%%%%%%%%%%%%%%%%%%%%%%%%%
\subsubsection*{$\bullet$ QCD parameters}
\nin
%%%%%%%%%%%%%%%%%%%%
%%%%%%%%%%%
%\subsubsection*{$\bullet$ QCD parameters}
\nin
The PT QCD parameters which appear in this analysis are $\alpha_s$, the charm and bottom quark masses $m_{c,b}$ (the light quark masses have been neglected). 
The non-perturbative condensates up to dimension 8 considered here  are the quark condensate $\langle\bar q q\rangle$, the two-gluon condensate $\langle g^2 G^2\rangle$, the mixed condensate $\langle g\bar q G q\rangle$, the four-quark condensate $\rho\langle\bar q q\rangle^2$, the three-gluon condensate $\langle g^3 G^3\rangle$, and the two-quark multiply two-gluon condensate $\rho\langle\bar q q\rangle\langle g^2 G^2\rangle$
  where $\rho$ indicates the deviation from the four-quark vacuum saturation. Their values are given in Table \ref{tab:parameter}. For the condensates, we shall use: 
\bea
 \langle\bar q q\rangle (\nu)=-\hat\mu^3_q\left(\rm{Log}\frac{\nu}{\Lambda}\right)^{-2/{\beta_1}}\\
\langle g\bar q G q\rangle(\nu)=-M^2_0\hat\mu^3_q\left(\rm{Log}\frac{\nu}{\Lambda}\right)^{-1/{3\beta_1}}
\eea   
where $\beta_1=-(1/2)(11-2n_f/3)$ is the first coefficient of the $\beta$ function, $\hat \mu_q$ the renormalization group invariant condensate and $\Lambda$ is the QCD scale.
%%%%%%%%%%%%%%%%%%%%%%%%%%%%%%%%
%\vspace*{-0.25cm}
\begin{table}[hbt]
\setlength{\tabcolsep}{1.2pc}
     {\small
\begin{tabular}{ll}
&\\
\hline
Parameters&Values.    \\
\hline
$\alpha_s(M_\tau)$&0.325(8)\\
$\Lambda(n_f=4)$& $(324\pm 15)$ MeV \\
$\Lambda(n_f=5)$& $(194\pm 10)$ MeV \\
$\bar m_c(m_c)$&$(1261\pm 24)$ MeV\\
$\bar m_b(m_b)$&$(4177\pm 22)$ MeV\\
$\hat \mu_q$&$(263\pm 7)$ MeV\\
$M_0^2$&$(0.8\pm 0.2)$ GeV$^2$\\
 $\langle\alpha_s G^2\rangle$&$(7\pm 2)\times 10^{-2}$ GeV$^4$\\
 $\langle g^3 G^3\rangle$&$(8.2\pm 2.0)$ GeV$^2 \times \langle \alpha_s G^2\rangle $\\
$\rho=\langle\bar q q\bar qq\rangle/\langle\bar q q\rangle^2$&$(2\pm 1)$ \\
\hline

\end{tabular}
}
\caption{\scriptsize    QCD input parameters (see e.g. \cite{SNB,SN14} and references therein).}
\label{tab:parameter}
\end{table}
%\vspace*{-0.2cm}
%%%%%%%%%%%%%%%%%%%%%%%%%%%%%%%%
%\vspace*{-0.25cm}
%%%%%%%%%%%%%%%%%%%%%%%%%%%%
\begin{figure}[hbt] 
\centerline {\hspace*{-6.5cm} a) }\vspace{-0.6cm}
\centerline{\includegraphics[width=5.2cm]{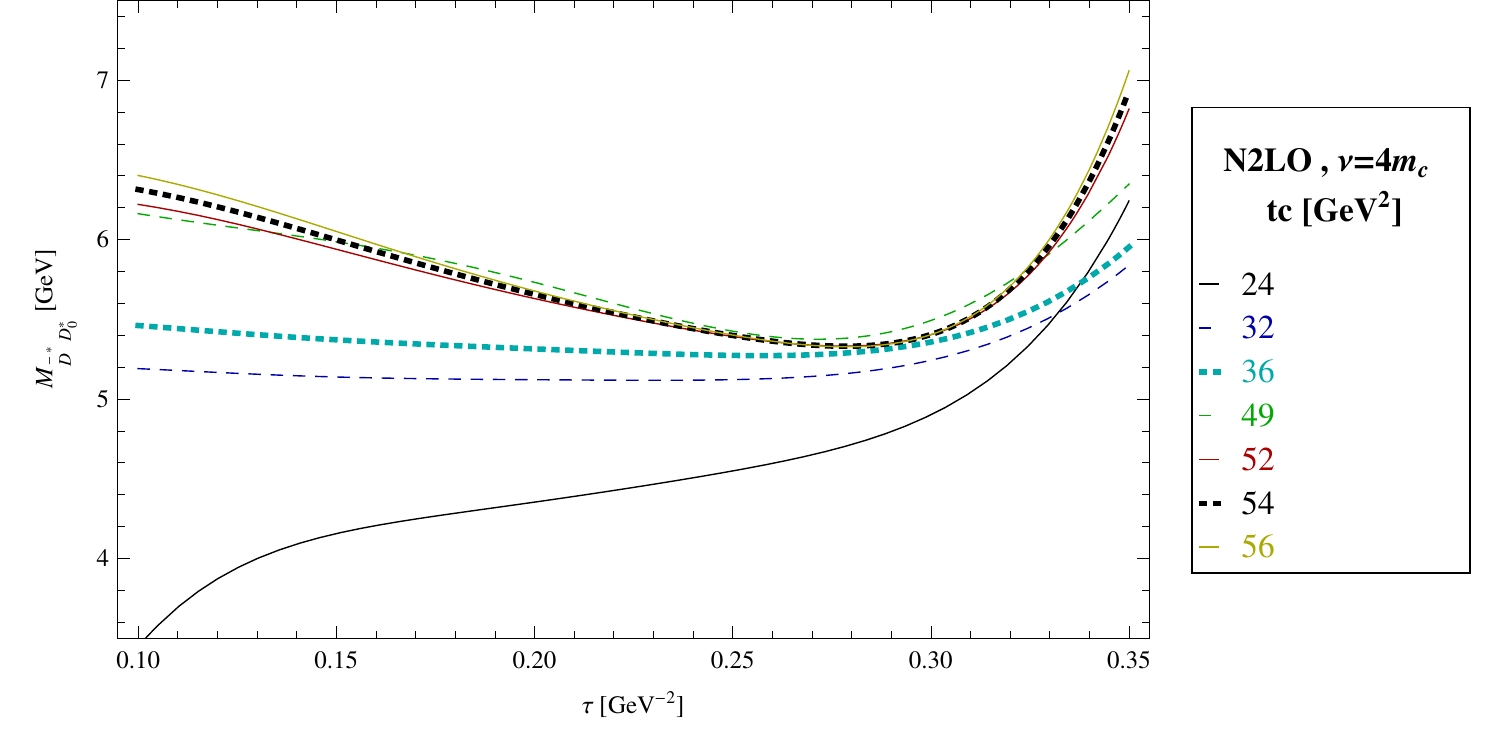}}
\centerline {\hspace*{-6.5cm} b) }\vspace{-0.6cm}
\centerline{\includegraphics[width=5.5cm]{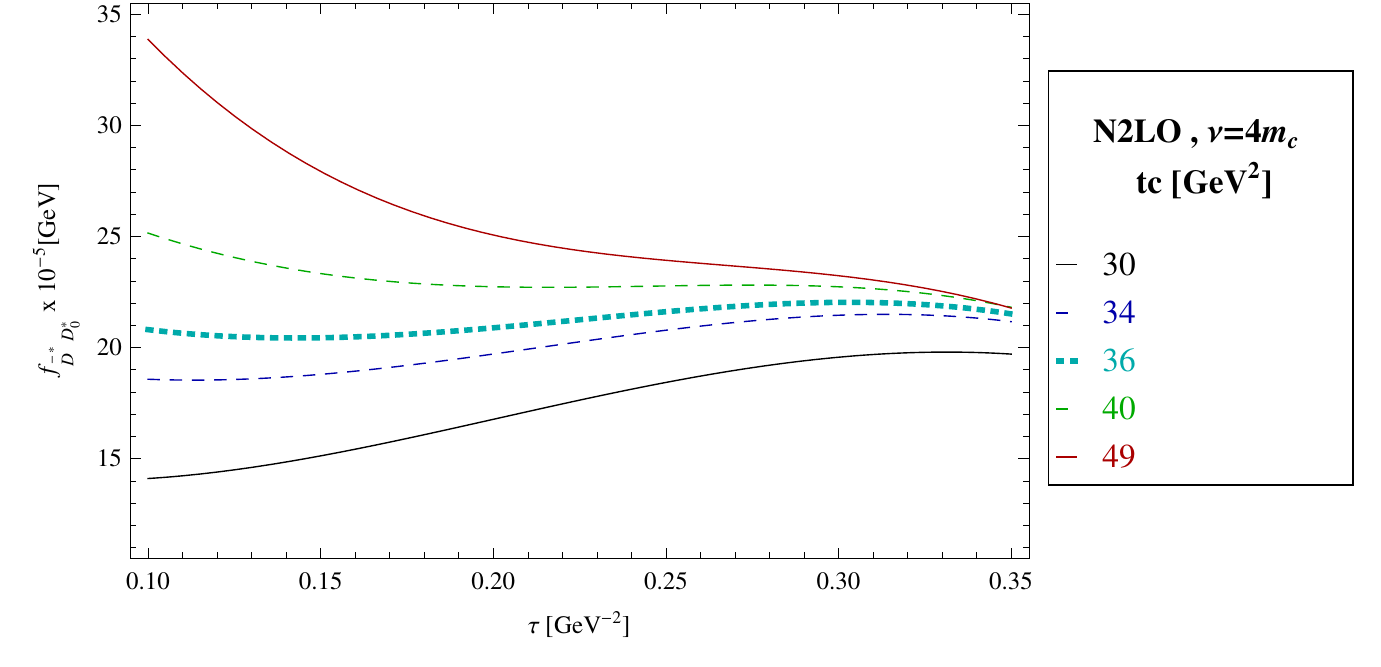}}
\caption{\scriptsize a) $\tau$-behaviour of the $M_{\bar{D}^*D^*_0}$ for different values of $t_c$ and for $\nu=4m_c$; the same as in a) but for the running coupling $f_{\bar{D}^*D^*_0}$.}
\label{fig1} 
\end{figure} 
\nin
%%%%%%%%%%%%%%%%%%%%
\begin{figure}[hbt] 
\centerline {\hspace*{-6.5cm} a) }\vspace{-0.6cm}
\centerline{\includegraphics[width=5.5cm]{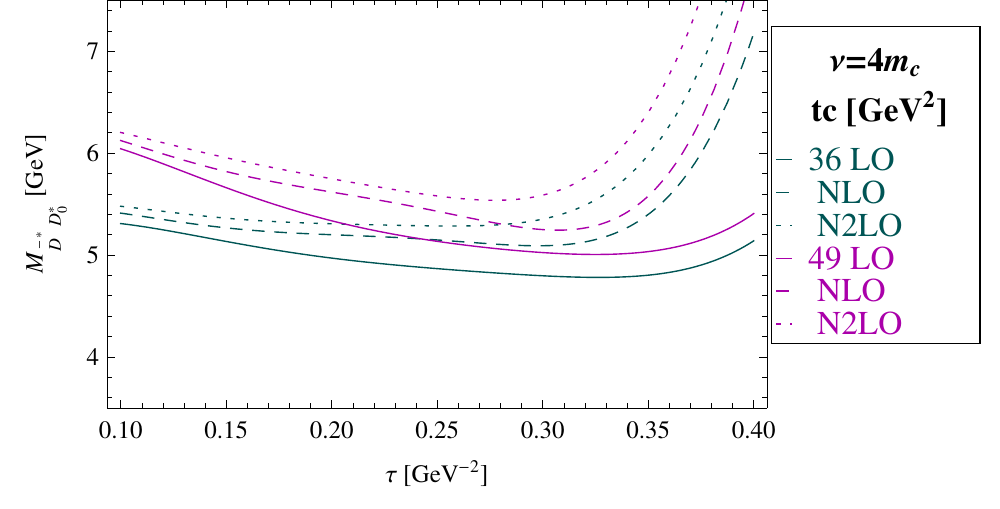}}
\centerline {\hspace*{-6.5cm} b) }\vspace{-0.6cm}
\centerline{\includegraphics[width=5.cm]{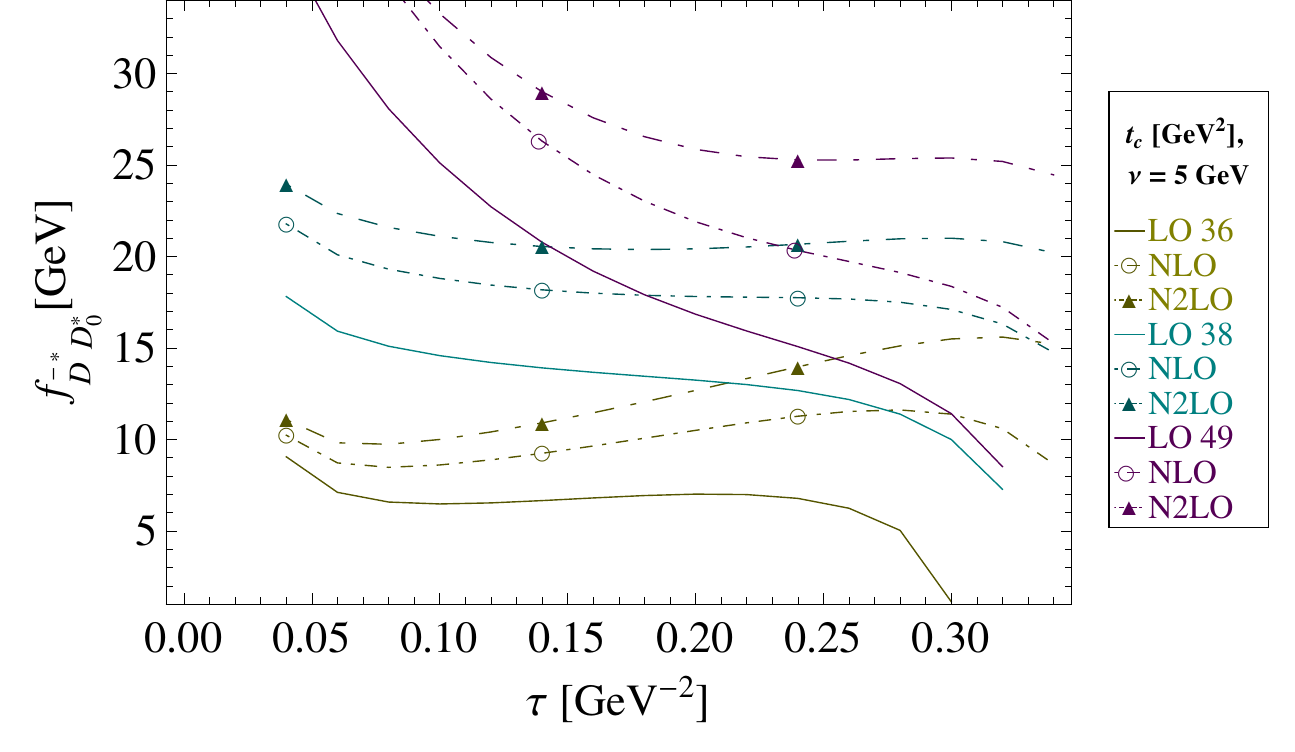}}
\caption{\scriptsize  a) $M_{\bar{D}^*D^*_0}$ mass versus $\tau$ for two extremal values of $t_c=36$ and 49 GeV$^2$ and for different truncation of the PT series; b) the as in a) but for but for the running coupling $f_{\bar{D}^*D_0}$.}
\label{fig3} 
\end{figure} 
\nin
%%%%%%%%%%%%%%%%%%%%%%%%%%%%%%%%%
\section{ $ \bar D^*D^*_0(1^{--})$ molecule mass and coupling }
%\vspace*{-0.2cm}
%%%%%%%%%%%%%%%%%%%%%%%%%%%%%%%%%
\subsubsection*{$\bullet$ $\tau$ and $t_c$ stabilities}
\nin
%%%%%%%%%%%%%%%%%%%%%%%%%%%%%%%%%%
We show  in Fig.~\ref{fig1}a the $\tau-$behaviour of the $M_{\bar{D}^*D^*_0}$  at N2LO of PT series for different values of $t_c$ at given subtraction $\nu=4m_c$ and in Fig.\,\ref{fig1}b the one of the coupling $f_{\bar{D}^*D^*_0}$, where one should note that the bilinear scalar heavy-light current acquires an anomalous dimension such that the decay constant runs as:
\beq
f_{\bar{D}^*D^*_0}(\nu)=\hat f_{\bar{D}^*D^*_0} \ga {\rm Log}{\nu\over \Lambda}\dr^{2/-\beta_1},
\eeq
where $\hat f_{\bar{D}^*D^*_0}$ is a scale invariant coupling.
One can see in these figures that the $\tau$ stability is obtained from $t_c= 32\sim 36 $ GeV$^2$ while $t_c$ stability is reached from $t_c=49$ GeV$^2$. We consider as optimal and conservative result the one obtained inside this region of $t_c$. 
%%%%%%%%%%%%%%%%%%%%%%%%%
\subsubsection*{$\bullet$ Convergence of the PT series} 
\nin
%
%\begin{table}
%%%%%%%%%%%%%%%%%%%%%%%
\nin
%%%%%%%%%%%%%%%%%%%%%%%
We show in the Fig.\,\ref{fig3}, the convergence of the expansion in PT. We can observe in Fig.\,\ref{fig3}a that, from the LO to the NLO, the mass increases by $+6.72\%$ and from LO $\oplus$ NLO to N2LO, by +4.42\%. These corrections indicate  that the LO  results obtained in\,\cite{AlbNie1,10tetra}, by using the value of the running $m_c$ mass, underestimates the molecule mass by about 11.14\%. \\
For the coupling, the $\alpha_s$ corrections are large. It increases about 50\% from LO to NLO and about 35\% from LO $\oplus$ NLO to N2LO, then to a total of 85\%. Neverthless, one can see that the PT series still converges though slowly. 
%%%%%%%%%%%%%%%%%%%%%%%%%%%% 
\subsubsection*{$\bullet$ $\nu$-stability}
\begin{figure}[hbt] 
\centerline {\hspace*{-6.5cm} a) }\vspace{-0.6cm}
\centerline{\includegraphics[width=5.5cm]{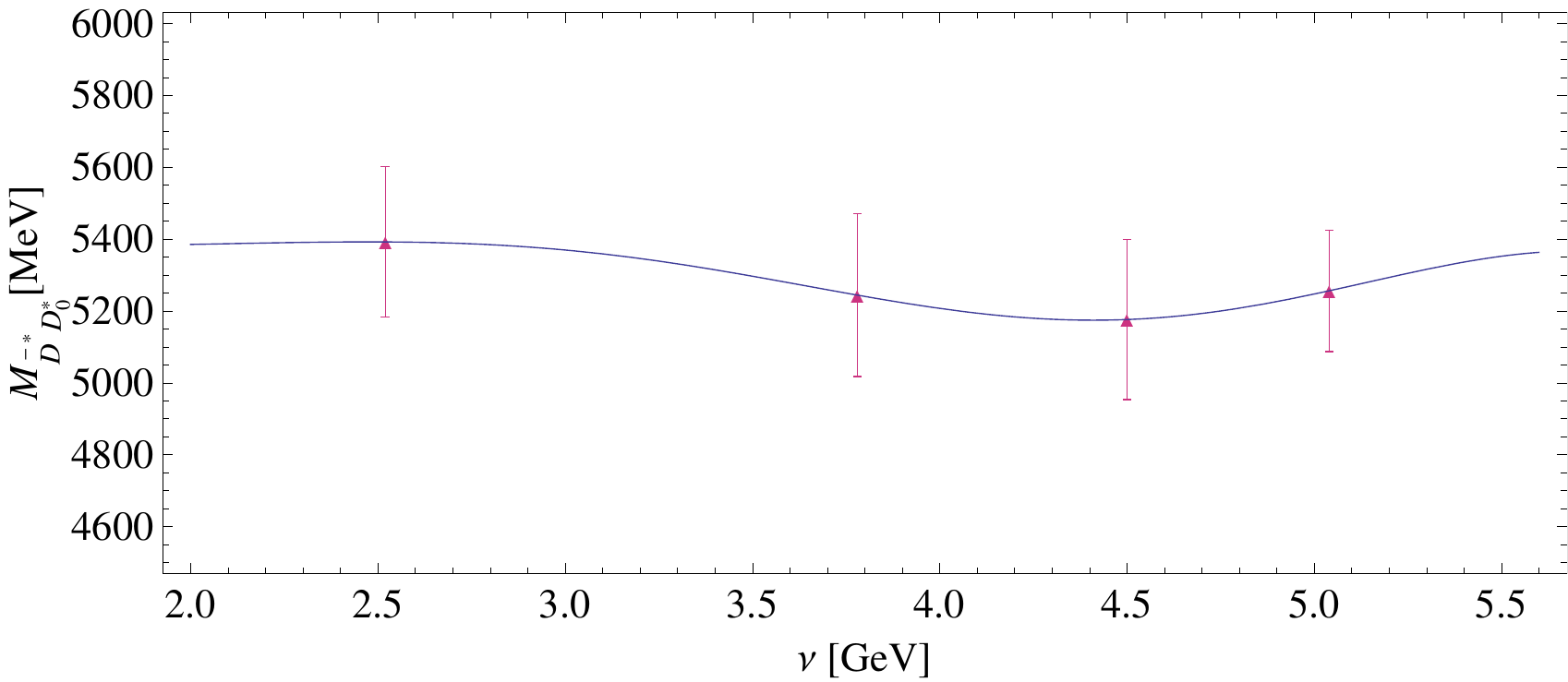}}
\centerline {\hspace*{-6.5cm} a) }\vspace{-0.6cm}
\centerline{\includegraphics[width=5.5cm]{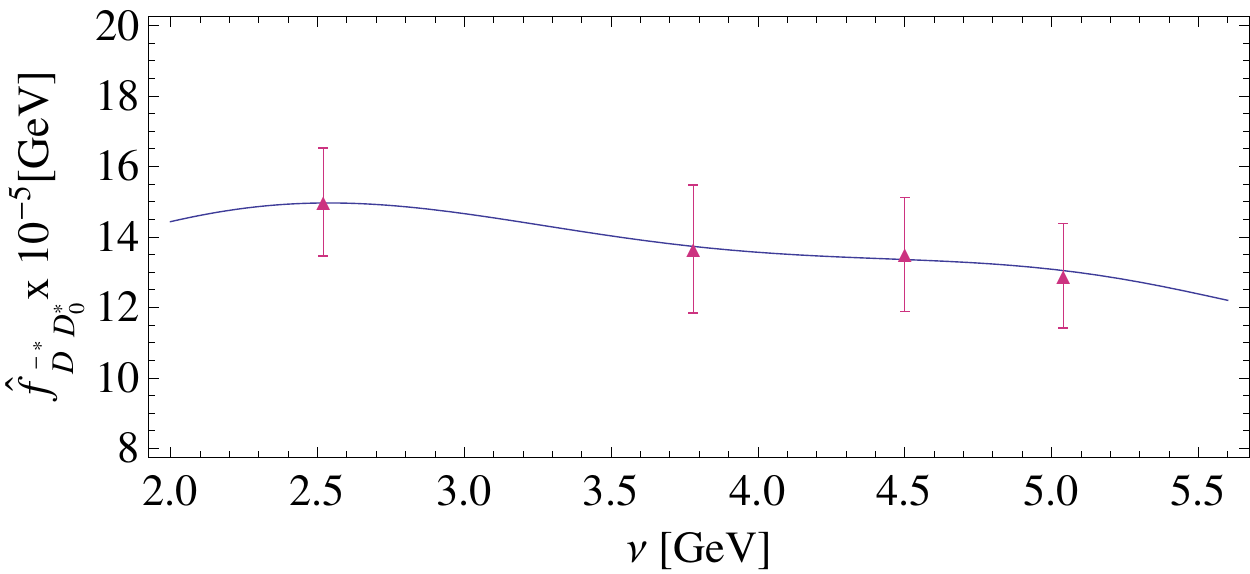}}
\caption{\scriptsize a) $\nu$-behaviour of the $M_{\bar{D}^*D^*_0}$; b)
the same as in a) but for $M_{\bar{D}^*D^*_0}$. }
\label{fig2} 
\end{figure}

\nin
%%%%%%%%%%%%%%%%%%%%%%%%%%
We study  in Fig.\,\ref{fig2}  the behaviour of the mass and of the invariant coupling $\hat  f_{\bar{D}^*D^*_0}$ in term of the scale $\nu$. One notices a good stability for $\nu$ ranging from 2.5 to 6 GeV where a minimum for the mass and an inflexion for the coupling occur for $\nu\approx 3m_c=3.8$ GeV which  the same value as the one in the $1^{++}$ channel. We obtain at N2LO:
\bea
M_{\bar{D}^*D^*_0}=5244(228) \textrm{ MeV}~,
\label{eq:mD}
\eea 
which is comparable with the LO result $M_{\bar{D}^*D^*_0}\simeq 5268(24)$ MeV obtained
by combining LSR and FESR\,\cite{10tetra}. For the coupling, we find:
\bea
 \hat f_{\bar{D}^*D^*_0}&=&(13.7\pm 0.9) \times 10^{-2} \textrm{MeV}~\lrar\nnb\\
 f_{\bar{D}^*D^*_0}(\nu)&=&(21.3 \pm 1.4) \times 10^{-2} \textrm{MeV}~,
  \label{eq:fD}
\eea
which we compare with the LO result $ f_{\bar{D}^*D^*_0}\approx 0.08$ MeV obtained in \cite{10tetra} for a $1^{--}$ four-quark state where the difference between the two values comes mainly from the large radiative corrections discussed before. 
%%%%%%%%%%%%%%%%%%%%%%%%%%%%
\nin
%%%%%%%%%%%%%%%%
\vspace*{-0.5cm}
%%%%%%%%%%%%%%%%%%%%%%%%%%
\section{$\bar{B}^*B_0(1^{--})$  mass and coupling}
%%%%%%%%%%%%%%%%%%%%%%%%%%%
\vspace*{-0.2cm}
%%%%%%%%%%%%%%%%%%%%%%%%%%%%%%%%%
\subsubsection*{$\bullet$ $\tau$ and $t_c$ stabilities}
\begin{figure}[hbt] 
\centerline {\hspace*{-6.5cm} a) }\vspace{-0.6cm}
\centerline{\includegraphics[width=5.2cm]{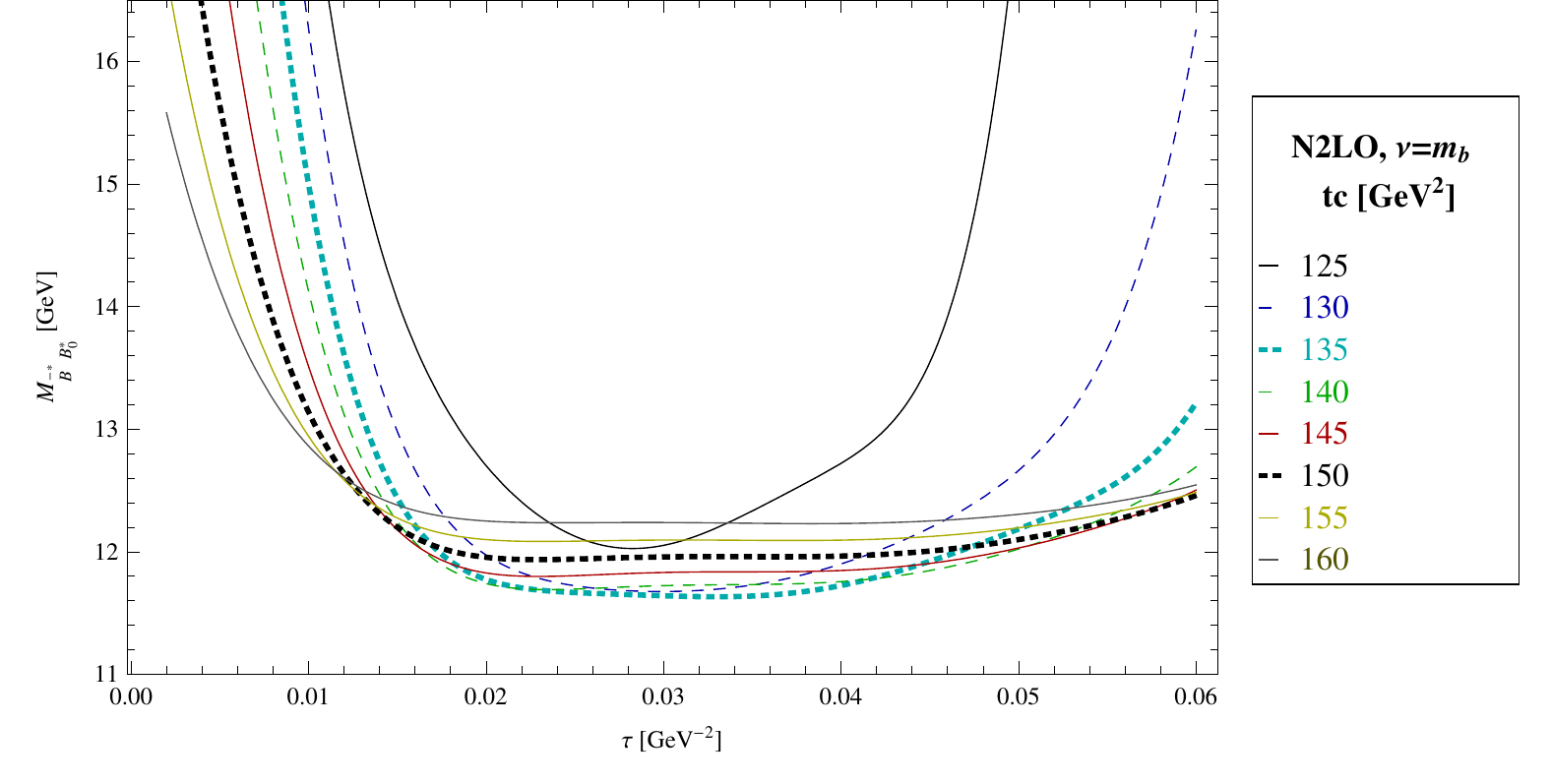}}
\centerline {\hspace*{-6.5cm} b) }\vspace{-0.6cm}
\centerline{\includegraphics[width=5.5cm]{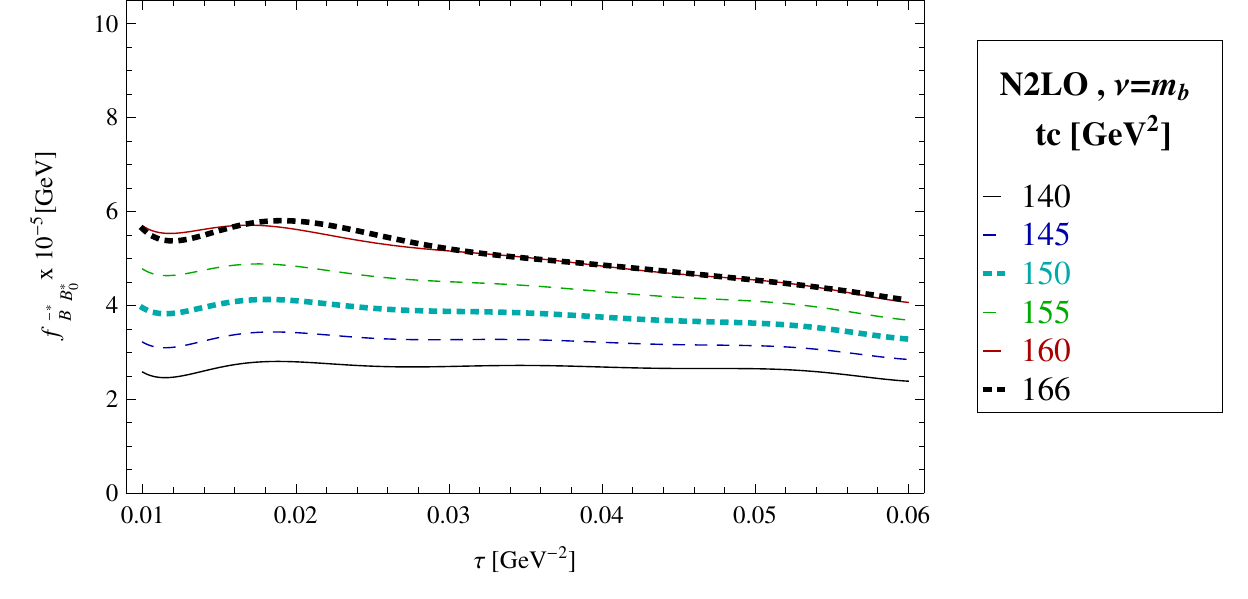}}
\caption{\scriptsize a) $\tau$-behaviour of the $M_{\bar{B}^*B^*_0}$ for different $t_c$ and for $\nu=m_b$; the same as in a) but for the running coupling $f_{\bar{B}^*B^*_0}$.}
\label{fig4} 
\end{figure}

\nin
%%%%%%%%%%%%%%%%%%%%%%%%%%%%%%%%%%
 We do the same analysis for the case of the beauty channel. The $\tau-$behavior of mass and running coupling is shown in the Fig.\,\ref{fig4} for  given value of $\nu$ and for different values of $t_c$. 
One can see that the stability in $\tau$ is between $t_c=135$ GeV$^2$ and $t_c=160$ GeV$^2$ where $t_c$ stability starts. We consider as optimal and conservative value of the mass and coupling the one obtained between these two extremal cases. 
%%%%%%%%%%%%%%%%%%%%%%%%%
\subsubsection*{$\bullet$ Convergence of the PT series} 
\begin{figure}[hbt] 
\centerline {\hspace*{-6.5cm} a) }\vspace{-0.6cm}
\centerline{\includegraphics[width=5.cm]{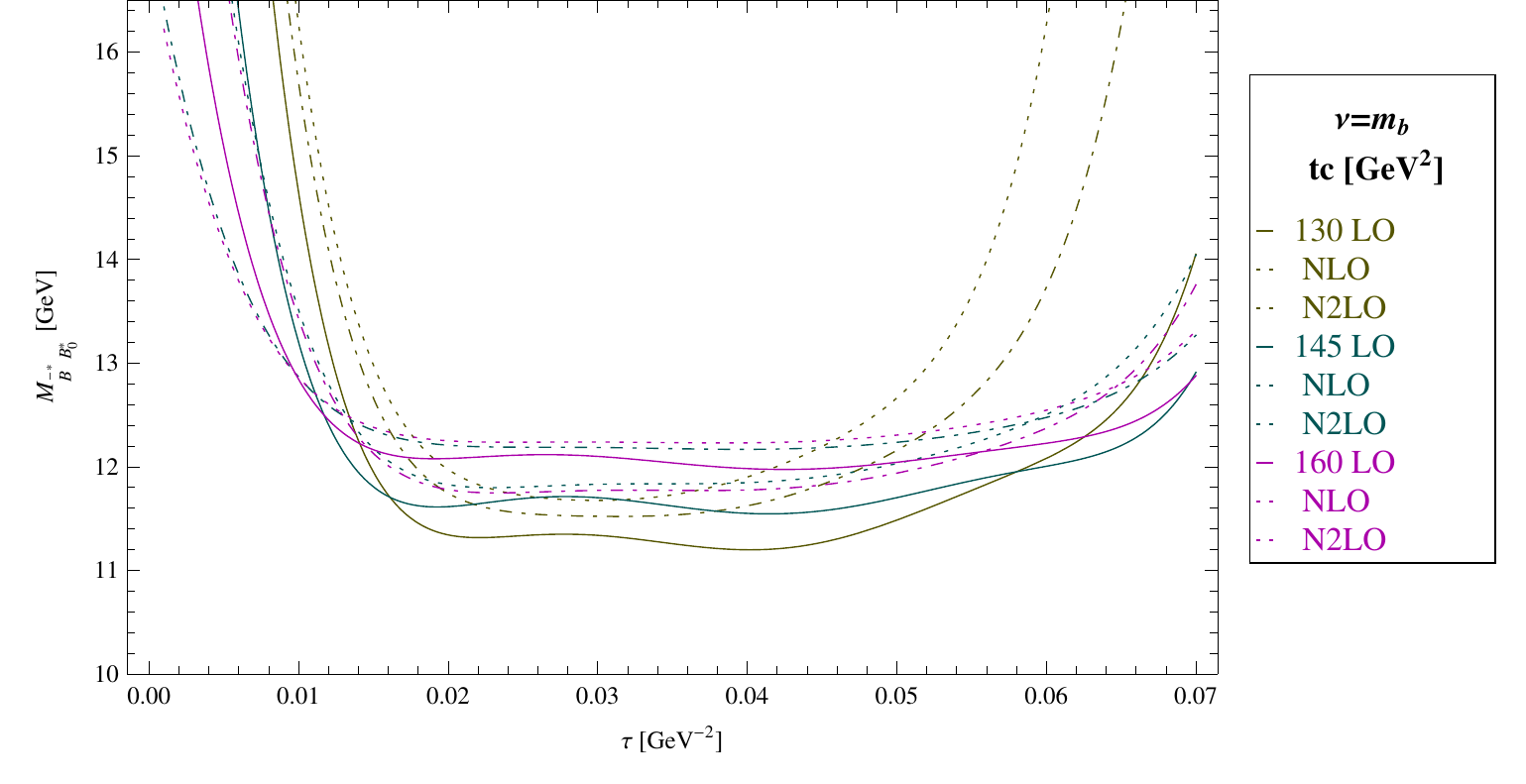}}
\centerline {\hspace*{-6.5cm} b) }\vspace{-0.6cm}
\centerline{\includegraphics[width=5.cm]{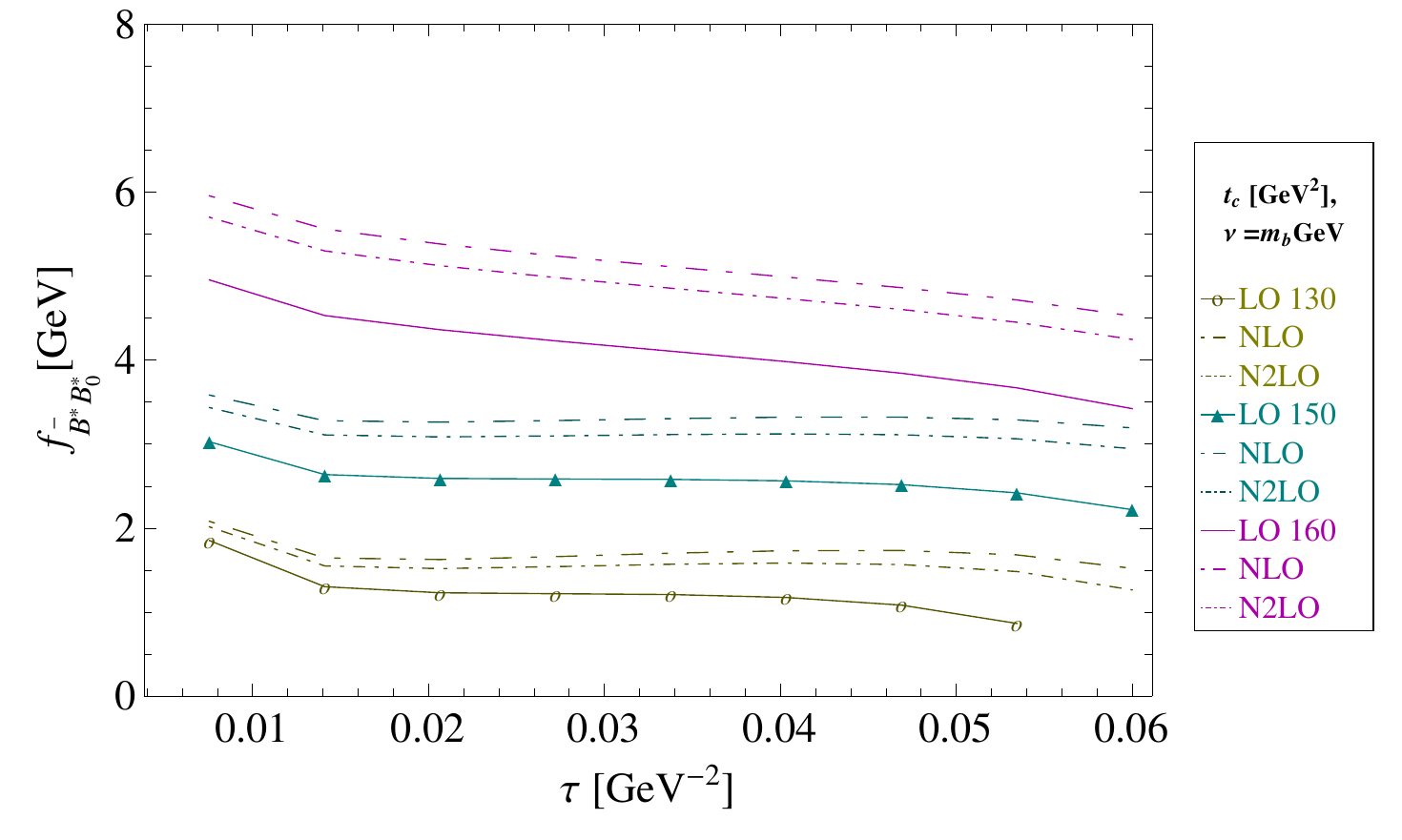}}
\caption{\scriptsize a) $M_{\bar{B}^*B^*_0}$ mass versus $\tau$ for two extremal values of $t_c=36$ and 49 GeV$^2$ and for different truncation of the PT series; b) the as in a) but for but for the running coupling $f_{\bar{B}^*B_0}$.}
\label{fig5} 
\end{figure}

\nin
%%%%%%%%%%%%%%%%%%%%%%%
We study in Fig.\,\ref{fig5} the convergence of the PT series. From LO to the NLO, the mass  increases by 4.9\% and from LO $\oplus$ NLO to N2LO it decreases by -2.98\% which shows small corrections and a good convergence of the PT series. For the coupling, the $\alpha_s$ correction increases the value by about 25\% while from LO $\oplus$ NLO to N2LO, its decreases by about -15\%. 
 %%%%%%%%%%%%%%%%%%%%%%%%%%%% 
\subsubsection*{$\bullet$ $\nu$-stability}
\begin{figure}[hbt] 
\centerline {\hspace*{-6.5cm} a) }\vspace{-0.6cm}
\centerline{\includegraphics[width=5.5cm]{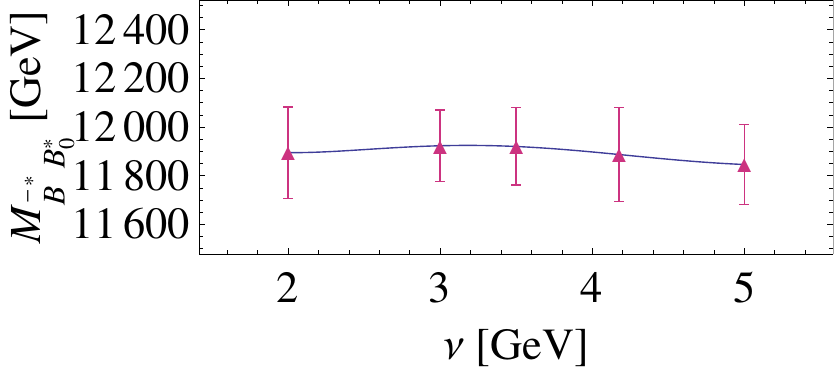}}
\centerline {\hspace*{-6.5cm} b) }\vspace{-0.6cm}
\centerline{\includegraphics[width=5.5cm]{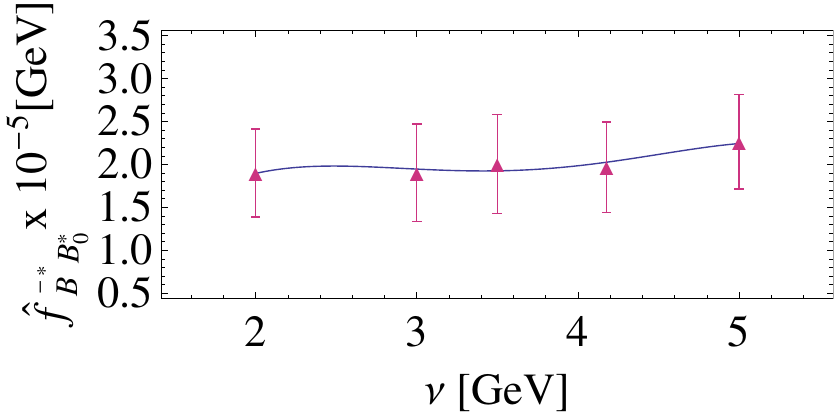}}
\caption{\scriptsize a) $\nu$-behaviour of the $M_{\bar{B}^*B^*_0}$; b)
the same as in a) but for $f_{\bar{B}^*B^*_0}$.}
\label{fig6} 
\end{figure}

\nin
%%%%%%%%%%%%%%%%%%%%%%%%%%
We study  in Fig.\,\ref{fig6}  the behaviour of the mass and of the invariant coupling $\hat  f_{\bar{B}^*B^*_0}$ in term of the scale $\nu$.
 Our final result for the mass comes from $\nu\approx 3.5$ GeV where stabilities are obtained both for the mass and for the couling. We obtain:
\bea
 M_{\bar{B}^*B^*_0} =11920(159)\textrm{ MeV}~.
 \label{eq:mB}
\eea
and:
\bea
\hat f_{\bar{B}^*B^*_0}&=&2.0 (0.59) \times 10^{-2} \textrm{ MeV} ~\lrar \nnb\\
f_{\bar{B}^*B^*_0}(\nu)&=&3.5 (1.02) \times 10^{-2} \textrm{ MeV}~.
 \label{eq:fB}
\eea
%%%%%%%%%%%%%%%%%%%%%%%%%%%%%%%%%%%%
\nin
%%%%%%%%%%%%%%%%%%%%%%%%%%%%%%%%
%%%%%%%%%%%%%%%%%%%%%%%%%%%%%%%%
%%%%%%%%%%%%%%%%%%%%%%%%%%%%%%%%%
\vspace*{-0.5cm}
%%%%%%%%%%%%%%%%%
\section{Conclusions}
\nin
%\vspace*{-0.1cm}
%%%%%%%%%%%%%%%%
We have studied the $\bar{D}^*D^*_0$ and $\bar{B}^*B^*_0$ $(1^{--})$ molecules states using QCD spectral sum rules to N2LO of PT series and including non-perturbative condensates up to dimension 8. We consider the present result as an improvement of the previous ones \cite{AlbNie1,10tetra}.\\
The mass of $\bar{D}^*D^*_0$ in Eq.\,(\ref{eq:mD}) at N2LO agrees within the error with the LO result \cite{AlbNie1,10tetra} due to the small radiative corrections in the ratio of moments, as expected from a general feature of this approach. The result shows that the $1^{--} $ experimental candidates Y(4260), Y(4360),  Y(4660)  are too light to be  pure $\bar{D}^*D^*_0$ molecule states. \\
The coupling $f_{\bar{D}^*D^*_0}$ obtained in Eq.\,(\ref{eq:fD}) is strongly affected by radiative corrections and using its  LO value for estimating some hadronic widths may lead to very inaccurate predictions. \\
For the beauty channel, our predictions in Eq.\,(\ref{eq:mB}) can be checked in different B factories.
Here, the radiative corrections are smaller for the coupling. \\
%%%%%%%%%%%%%%%%%%%%%%%%%%%
\vspace*{-0.5cm}
\section*{Acknowledgements}
\nin
%%%%%%%%%%%%%%%%
F.F. and A.R. would like to thank the CNRS for supporting the travel and living expenses and the LUPM-Montpellier for hospitality. We also thank R.M. Albuquerque for many helpful discussions.
%%%%%%%%%%%%%%%%%%%%%
\input{bib_fenosoa}

\end{document}

%% file: bib_fenosoa.tex
%%%%%%%%%%%%%%%

%% file: qcd-fenosoa.bbl
\begin{thebibliography}{999}
\vspace*{-0.25cm}
%%%%%%%%%%%%%%%%%%%
% QSSR SVZ & REVIEW
%%%%%%%%%%%%%%%%%%%
\bibitem{BELLE}B. Aubert et al. (BaBar Collaboration), {\it Phys. Rev. Lett.}
{\bf 95} (2005) 142001;  Q. He et al.(CLEO Collaboration), {\it Phys. Rev.} {\bf D74} 
(2006) 091104(R);  C. Z. Yuan et al. (Belle Collaboration), {\it Phys. Rev. Lett.} {\bf 99}
 (2007) 182004; T. E. Coan et al. (CLEO Collaboration), {\it Phys. Rev. Lett.} {\bf 96} (2006) 162003;
 J. P. Lees et al. (BaBar Collaboration), {\it Phys. Rev.} {\bf D86} (2012) 051102(R).
 
\bibitem{REVMOL}For reviews, see e.g.: F. S. Navarra,
M. Nielsen, S. H. Lee, {\it Phys. Rep.}  {\bf 497} (2010) 4; S. L. Zhu, {\it Int. J. Mod. Phys.}  {\bf E17} (2008); E.Swanson,  {\it Phys. Rep.} {\bf 429} (2006) 243; N. Brambilla et al., {\it Eur. Phys. J}  {\bf C71} (2011) 1534.

 \bibitem{10tetra} R.M. Albuquerque, F. Fanomezana, S. Narison, A. Rabemananjara, {\it Phys. Lett.} {\bf B 715} (2012) 129.
\bibitem{SVZ} M.A. Shifman, A.I. and Vainshtein and V.I. Zakharov,
Nucl. Phys. {\bf B147} (1979) 385 .
\bibitem{SNB} For a review and references to original works, see
e.g., S. 
Narison, {\it QCD as a theory of hadrons,
Cambridge Monogr. Part. Phys. Nucl. Phys. Cosmol.} {\bf 17} 
(2002) 1-778
[hep-h/0205006]. 
\bibitem{PICH} A. Pich and E. de Rafael, {\it Phys. Lett.}{\bf B158} (1985) 477.
\bibitem{SNPIVO} S. Narison and A. Pivovarov, {\it Phys. Lett.} {\bf B327} (1994) 341.
\bibitem{SNFB12} S. Narison, {\it Phys. Lett.} {\bf B718} (2013) 1321.
\bibitem{SNFBST14}S. Narison, arXiv:1404.6642 [hep-ph].
\bibitem{GENERALIS}S.C. Generalis, Ph.D. thesis, Open Univ. report, OUT-4102-13 (1982), unpublished.
\bibitem{CHET}K.G. Chetyrkin and M. Steinhauser, {\it Phys. Lett.} {\bf B 502} (2001) 104; hep-ph/0108017.
\bibitem{SPEC1} S. Narison, {\it Phys. Lett.} {\bf B197} (1987) 405; 
%S. Narison, 
{\it Phys. Lett.}{\bf B 216}
(1989) 191.
\bibitem{SPEC2} K.G. Chetyrkin and M. Steinhauser, {\it Nucl. Phys.} {\bf B 573} (2000) 617; K.
Melnikov and T. van Ritbergen, hep-ph/9912391.
\bibitem{AlbNie1} R. M. Albuquerque and M. Nielsen, {\it Nucl. Phys.}{ \bf A815} (2009) 53;
Erratum-ibid. {\bf A857} (2011) 48 and private notes from R. Albuquerque.
\bibitem{SN14} S. Narison, arXiv:1409.8148 (talk given at this conference).
 \end{thebibliography}
